\documentclass{iau}
\usepackage[dvips]{graphicx}
\usepackage{subfig}
\usepackage{upmath}

\title[Performances tests on the SPHERE-IFS] 
{Performances tests on the SPHERE-IFS}

\author[Dino Mesa et al.]   
{Dino Mesa$^1$, Raffaele Gratton$^1$, Riccardo U. Claudi$^1$, Silvano Desidera$^1$,
Enrico Giro$^1$, Alice Zurlo$^{1,2}$, Anne Costille$^2$, Arthur Vigan$^2$, Claire Moutou$^2$,
Jean-Luc Beuzit$^3$, Kjetil Dohlen$^2$, Markus Feldt$^4$, David Mouillet$^3$, 
Jean-Francois Sauvage$^3$, Markus Kasper$^5$
 \and Jacopo Antichi$^6$}

\affiliation{$^1$INAF - Osservatorio Astronomico di Padova - Vicolo dell'Osservatorio 5, I-35122, Padova, Italy \\[\affilskip]
$^2$Aix Marseille Universite, CNRS, LAM (Laboratoire d'Astrophysique de Marseille) UMR 7326, 13388, Marseille, France \\[\affilskip]
$^3$Institut de Planetologie et d'Astrophysique de Grenoble, UJF, CNRS, 414 rue de la piscine, 38400, Saint Martin d'Heres, France \\[\affilskip]
$^4$Max-Planck-Institut fur Astronomie, Konigstuhl 17, D-69117 Heidelberg, Germany \\[\affilskip]
$^5$European Southern Observatory, Karl Schwarzschild Strasse, 2, D-85748 Garching bei Munchen, Germany \\[\affilskip]
$^6$INAF - Osservatorio Astrofisico di Arcetri - L.go E. Fermi 5, I-50125 Firenze, Italy}

\pubyear{2013}
\volume{299}  
\pagerange{119--126}
\jname{Exploring the formation and evolution of planetary systems}
\editors{A.C. Editor, B.D. Editor \& C.E. Editor, eds.}
\begin{document}

\maketitle

\begin{abstract}
Until now, just a few extrasolar planets (~30 out of 860) have been found through the direct 
imaging method. This number should greatly improve when the next generation of High Contrast 
Instruments like Gemini Planet Imager (GPI) at Gemini South Telescope or SPHERE at VLT will 
became operative at the end of this year. In particular, the Integral Field Spectrograph (IFS), 
one of the SPHERE subsystems, should allow a first characterization of the spectral type of the 
found extrasolar planets.
Here we present the results of the last performance tests that we have done on the IFS 
instrument at the Institut de Planetologie et d'Astrophysique de Grenoble (IPAG) in condition 
as similar as possible to the ones that we will find at the telescope. We have found that we 
should be able to reach contrast down to 5x10$^{-7}$ and make astrometry at sub-mas level with 
the instrument in the actual conditions. A number of critical issues have been identified. The 
resolution of these problems could allow to further improve the performance of the instrument.
\end{abstract}

\firstsection 
\section{Introduction}

Imaging of extrasolar planets is a very challenging goal because of the very large 
luminosity contrast (of the order 10$^{-6}$ for young giant planets and of the order of 
10$^{-8}$-10$^{-10}$ for old giant and rocky planets) and the small angular separation (few 
tenths of arcsec for a planet at $\sim$10 AU around a star at some tens of pc) between the host 
star and the companion objects. 
However, a number of different project are either now running (e.g. Project 1640 at the 5 m 
Palomar Telescope - see~\cite[Crepp et al. 2011]{Crepp2011}) or are going to begin like the 
Gemini Planet Imager (GPI) at 
the Gemini South Telescope~(\cite[Macintosh et al. 2006]{Macintosh2006}) or SPHERE at the ESO 
Very Large Telescope~(\cite[Beuzit et al. 2006]{Beuzit2006}).
This last instrument, in particular, includes three scientific channels that are a 
differential imager and dual band polarimeter called IRDIS operating in the near infrared 
between the Y and the Ks band (\cite[Dohlen et al. 2008]{Dohlen2008}), a polarimeter called 
ZIMPOL that will look for old planets at visible wavelengths (\cite[Thalmann et al. 2008]
{Thalmann2008}) and an Integral Field Spectrograph (IFS) operating in the near infrared between
the Y and the H band (\cite[Claudi et al. 2008]{Claudi2008}). In the next paragraphs we will
present the results of the laboratory tests on the IFS. 

\section{Test description}

Tests on the IFS instrument were held in January and February 2013 at the \textit{Institut de 
Planetologie et d'Astrophysique de Grenoble} (IPAG) facility with the aim to validate 
functionality of the science and calibration templates and to preliminary estimate the 
performances of the instrument. The tests were performed both in the YJ (0.95$\div$1.35 
micron) and in YH (0.95$\div$1.65 micron) mode using the appropriate combination of Lyot 
coronagraph and apodized mask.

Data were then reduced exploiting the Data Reduction and Handling (DRH) software that 
allows to perform all the required calibrations and the speckle subtraction procedure through 
the spectral deconvolution (SD) method (\cite[Sparks \& Ford 2002]{Sparks2002}). A 
further speckle suppression can be obtained applying angular differential imaging (ADI)
(\cite[Marois et al. 2006]{Marois2006}). Given that we do cannot perform any rotation of the 
field of view during our tests, we can just perform a simulation of the method so that our 
results have to be regarded as just an estimation of the contrast that we will be able to get. 

\section{Results}

\begin{figure}[b]
\begin{center}
 \subfloat{\includegraphics[width=6.0cm]{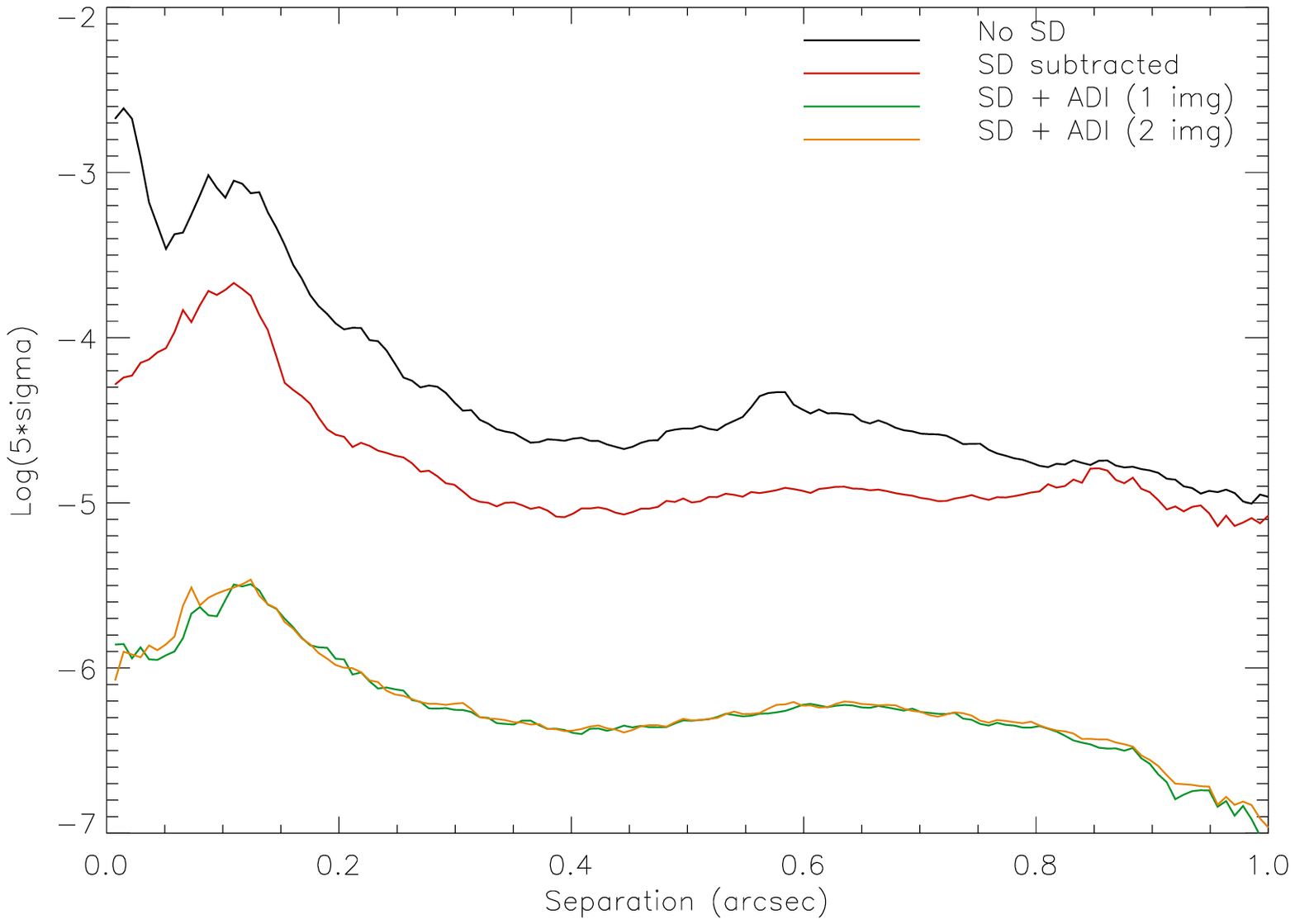}} \quad
 \subfloat{\includegraphics[width=6.0cm]{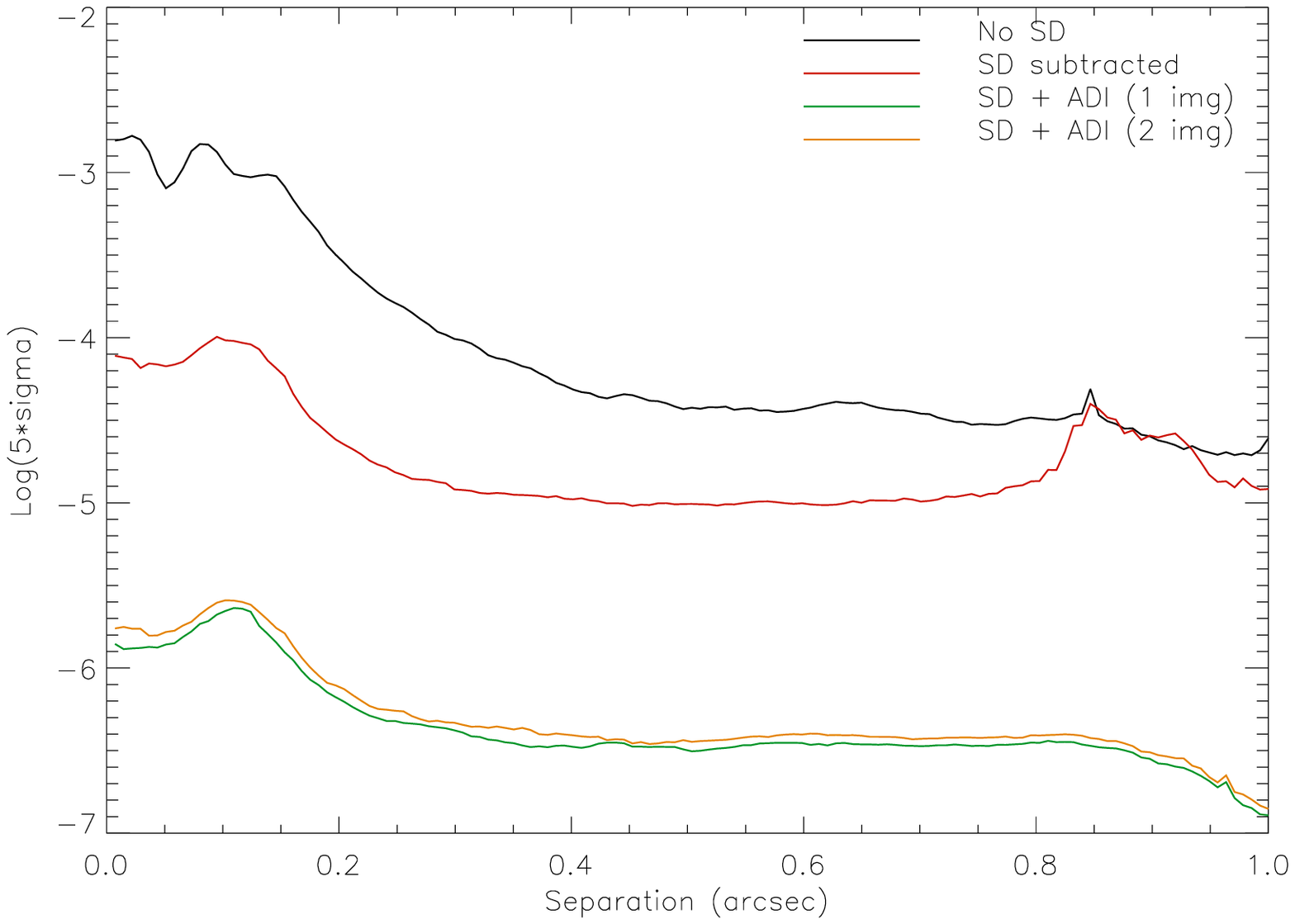}} 
 \caption{Contrast plot for IFS operating in the YJ-mode (left panel) and in the YH-mode (right
 panel). }
   \label{fig2}
\end{center}
\end{figure}

In Figure~\ref{fig2} we display the 5$\sigma$ contrast plot that we can get for the IFS 
operating both in the YJ-mode (left panel) and in the YH-mode (right panel). A contrast better than 10$^{-6}$ can be obtained for both the modes appropriately combining SD and ADI. To 
further confirm this results we add a number of simulated planets to the raw data at different separations and with luminosity contrast of 10$^{-5}$ and 10$^{-6}$ and reduced these data following the same procedure. All simulated planets are visible with a S/N greater than 5.

\end{document}